\documentclass[prl,superscriptaddress,amsmath,amssymb,aps,twocolumn,showpacs]{revtex4}
\usepackage[]{inputenc}
\usepackage[dvips]{graphicx}
\usepackage{color}
\usepackage{bm}
\usepackage{mathpazo}
\usepackage{slashed}
\usepackage[ps2pdf,bookmarks=true,colorlinks,linkcolor=blue,urlcolor=green,citecolor=red]{hyperref}

\renewcommand{\part}{{\rm part}}

\newcommand{\be}{\begin{equation}}
\newcommand{\ee}{\end{equation}}
\newcommand{\bear}{\begin{eqnarray}}
\newcommand{\eear}{\end{eqnarray}}
\newcommand{\ba}{\begin{array}}
\newcommand{\ea}{\end{array}}

\begin{document}

\title{A quarksonic matter at high isospin density}

\author{Gaoqing Cao}
\affiliation{Physics Department and Center for Particle Physics and Field Theory, Fudan University, Shanghai 200433, China.}
\author{Lianyi He}
\affiliation{Department of Physics, Tsinghua University and Collaborative Innovation Center of Quantum Matter, Beijing 100084, China.}
\author{Xu-Guang Huang}
\affiliation{Physics Department and Center for Particle Physics and Field Theory, Fudan University, Shanghai 200433, China.}

\date{\today}

\begin{abstract}
Analogous to the quarkyonic matter at high baryon density in which the quark Fermi seas and the baryonic excitations coexist, it is argued that a "quarksonic matter" phase appears at high isospin density where the quark (antiquark) Fermi seas and the mesonic excitations coexist. We explore this phase in detail in both large $N_c$ and asymptotically free limits: In large $N_c$ limit, we sketch a phase diagram for the quarksonic matter. In the asymptotically free limit, we study the pion superfluidity and thermodynamics of the quarksonic matter by using both perturbative calculations and effective model.
\end{abstract}

\maketitle

\section{Introduction}
In 1932, by direct analogy with the ordinary spin, Heisenberg introduced the concept of isospin to describe the symmetry between proton and neutron under strong interaction~\cite{Heisenberg:1932dw}. Then, after the discovery of the substructure of hadrons~\cite{GellMann:1964nj}, the isospin was naturally transferred to denote the two lightest flavors of quarks, the $u$ and $d$ quarks. Now, the isospin is a widely used concept in nuclear and particle physics. The systems with nonzero isospin densities span a large range of nuclear medium from the normal nuclei near drop lines to the neutron stars~\cite{Li:2008gp}. Novel phenomena can happen at high isospin density. For example, it is well known that pion superfluidity can occur at large isospin chemical potential $\mu_I$ (or density)~\cite{Son:2000xc,Son:2000by,He:2005nk,Sun:2007fc,Cohen:2015soa,Brauner:2016lkh} and such superfluidity experiences a smooth crossover from the Bose-Einstein condensation of pions to Bardeen-Cooper-Schrieffer quark-antiquark pairing (BEC-BCS crossover) with increasing $\mu_I$. Moreover, as there is no sign problem at finite $\mu_I$, these results were also studied and verified by lattice quantum chromodynamics (LQCD) simulations~\cite{Kogut:2002zg,Kogut:2004zg,Bali:2016nqn}.

Based on large $N_c$ analysis, McLerran and Pisarski argued that, at large baryon chemical potential $\mu_B$, a new phase of dense matter can emerge in which the chiral symmetry is restored while the system remains confined at low temperature~\cite{McLerran:2007qj}. They call such a new phase of matter the ``quarkyonic matter" which reflects the coexistence of quarks and baryons: the thermodynamic properties of the system are dominated by quark Fermi seas but the elementary fermionic excitations are confined baryons near the Fermi surfaces. Interestingly, recent LQCD simulation showed that even at $N_c=2$, the notion of quarkyonic matter is serviceable~\cite{Braguta:2016cpw}. In this case, the LQCD found that the deconfinement never happens at zero temperature when $\mu_B$ increases; instead, the system at large $\mu_B$ is characterized by quarks always residing in Fermi seas and diquark condensate near the Fermi surface.

\begin{figure}[!htb]
\begin{center}
\includegraphics[width=5cm,height=6cm]{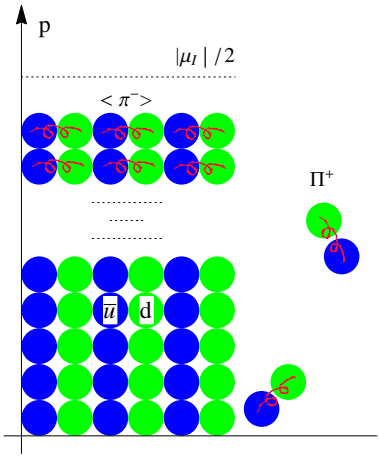}
\caption{The particle distributions in quarksonic matter at low temperature and large isospin density where the highest dotted line is the Fermi surface $p_F=|\mu_I|/2$ associated with isospin density, $\langle\pi^-\rangle$ denotes the pion condensate near Fermi surface and $\Pi^+$ denotes the Goldstone boson related to $I_3$ flavor symmetry breaking. The red spirals represent gluons.}\label{quarkson}
\end{center}
\end{figure}

As well known, that the phase structure of the two-color QCD (QC$_2$D) at finite baryon chemical potential $\mu_B$ is very similar with that of real QCD at finite isospin chemical potential $\mu_I$~\cite{Sun:2007fc,Brauner:2009gu}. In fact, they are identical in large $N_c$ limit~\cite{Hanada:2011jb}. Thus, one can anticipate that the QCD matter at large isospin density is also quarkyonic matter-like but with a Fermi sea comprising quark (antiquark) isospin charge and elementary excitation near the Fermi surface the mixing pions some isospin component of which condenses at low temperature and leads to superfluidity. To be more specific, we call this phase of matter the ``quarksonic matter". Its physical picture is illustrated in Fig. \ref{quarkson} where the case $\mu_I<0$ is considered: Near the Fermi surface of isospin charge, $\bar u$ antiquarks and $d$ quarks pair with each other to form pion condensate $\langle\pi^-\rangle$ at low temperature. For large enough $|\mu_I|$, the pressure and $|\mu_I|$ related quantities such as isospin density and compressibility can well be described by nearly free fermion gases of $\bar{u}$ antiquarks and $d$ quarks; but the thermally excited quantities such as entropy and heat capacity can be well described by Goldstone excitation $\Pi^+$ at low temperature~\cite{Son:2000by,Son:2000xc}. As temperature increases, the deconfinement phase transition happens at a certain critical temperature $T_d$ above which gluons and then quarks (antiquarks) become active thermal excitations.

In this work, we will revisit the quarksonic matter and its properties in both large $N_c$ and asymptotically free limits. In large $N_c$ limit, we will map out a phase diagram in the temperature and isospin chemical potential plane along with discussions similar with those of quarkyonic matter in Ref.~\cite{McLerran:2007qj}. Some comments on the real QCD phase diagram will also be given right after that. In asymptotically free limit, we will first derive the coefficient $b_\pi$ (see below for the definition) involved in the expression of the pion condensate up to sub-leading order in QCD coupling constant following Refs.~\cite{Pisarski:1999tv,Wang:2001aq} and then perform further numerical calculations by utilizing the effective pure gluodynamics formalism developed in Ref.~\cite{Cohen:2015soa} to study the characters of the new quarksonic matter phase.

\section{Large $N_c$ limit and phase diagram}
\begin{figure}[!htb]
\begin{center}
\includegraphics[width=8cm]{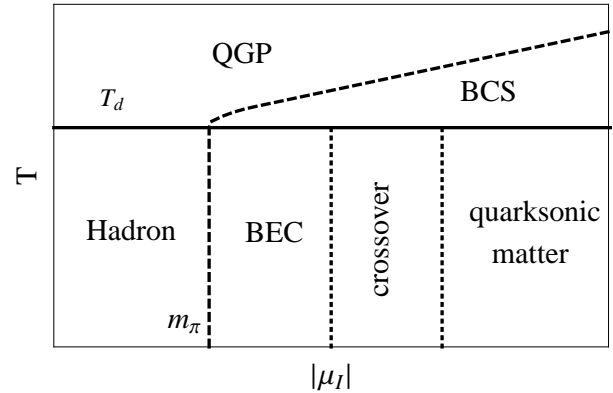}
\caption{The illustration of $T-|\mu_I|$ phase diagram in large $N_c$ limit. The solid, dashed and dotted lines correspond to the first-order transition, second-order transition, and crossover, respectively.}\label{phase diagram}
\end{center}
\end{figure}

The real QCD system with $N_c=3$ is strongly coupled at low energy and thus is very difficult to handle. But if one treats $N_c$ as a free parameter and lets it go to infinity while keeps $\lambda=g^2 N_c$ fixed (the 't Hooft limit)~\cite{'tHooft:1973jz}, one can greatly simplify the problem and obtain important insights to the original problem. So we will first consider the phase diagram in $T-|\mu_I|$ plane at large $N_c$ limit which is illustrated in Fig.~\ref{phase diagram}. Two important points should be emphasized afore: First, as chiral symmetry restoration is usually consistent with deconfinement transition, only BCS state is expected to exist above deconfinement temperature $T_d$. Second, due to volume independence in large-$N_c$ QCD-like gauge theories~\cite{Kovtun:2007py}, all the transition lines below $T_d$ are temperature independent as had been pointed out in Ref.~\cite{Hidaka:2011jj}.

At large $N_c$ limit, the quark loops ($\sim N_c^1$) are strongly suppressed compared to gluon loops ($\sim N_c^2$)~\cite{'tHooft:1973jz} which is responsible for the confining dynamics. As a result, the deconfinement temperature is independent of $|\mu_I|$, $T_d\sim \Lambda_{QCD}$, where $\Lambda_{QCD}$ is the QCD confinement scale which is assumed to be $\sim N_c^0$. At zero temperature, the pion superfluidity happens when $|\mu_I|$ is larger than the pion mass in vacuum $m_\pi$~\cite{Son:2000xc,Son:2000by} which is $\sim N_c^0$; thus the critical isospin chemical potential is $|\mu_I^c|\sim N_c^0$.  Right above $|\mu_I^c|$, the system is a dilute Bose gas of pions and the superfluidity is due to the BEC of pions. In this region, the pressure is of order $\sim N_c^0$ as in hadronic phase, because the isospin density ($\sim N_c^1$) almost all contributes to the BEC which doesn't affect the thermodynamic properties. With increasing $\mu_I$, the system begins to transform from BEC of pions to BCS superfluidity of quarks and antiquarks. There is no symmetry breaking occurs during such a BEC-BCS crossover and thus no critical $|\mu_I|$ to set this crossover. We mark the start point of the BEC-BCS crossover as the $|\mu_I|$ at which the in-medium quark mass drops to zero $m_q(|\mu_I|)-|\mu_I|/2=0$. Such determined $|\mu_I|$ is also independent of $N_c$. Eventually, when $|\mu_I|$ is large enough, the pressure is dominated by quark (antiquark) Fermi seas and is of order $\sim N_c^1$., then we enter the quarksonic matter phase. Because the interaction among pions are suppressed at large $N_c$ limit, we expect the threshold $|\mu_I|$ for this phase is also $\sim N_c^0$. The pion condensate or the quark-antiquark condensate in general increases with $|\mu_I|$ and so does the critical temperature $T_\pi$ for pion superfluidity. Thus for large enough $|\mu_I|$, $T_\pi$ will eventually be larger than $T_d$ leaving the region in between being occupied by deconfined pion superfluid. Above $T_\pi$, the system is a quark-gluon plasma (QGP) with pressure $\sim N_c^2$.

For real QCD, the topological structure of the $T-|\mu_I|$ phase diagram is expected to be similar with the one in large $N_c$ limit. However, some detailed structures can be altered. Combining the results from perturbative discussions~\cite{Son:2000xc,Son:2000by}, LQCD calculations~\cite{Kogut:2002zg,Kogut:2004zg}, and also model studies~\cite{He:2005nk,Sun:2007fc}, one expects $T_d$ to decrease with increasing $|\mu_I|$ in the whole region (as the confinement scale is decreasing at large $|\mu_I|$) which indicates shrink of the quarksonic matter region. The melting temperature of the pion superfluidity keeps increasing with  $|\mu_I|$ and the transition is still second order, but the deconfinement transition was argued to change from crossover to first order with a new critical end point in the BEC-BCS crossover region~\cite{Son:2000xc,Son:2000by,Kogut:2004zg}.

\section{Asymptotically free limit}
Thanks to the asymptotic freedom of QCD, we can perform perturbative calculation when $|\mu_I|$ is very large.
Previously, similar calculations have already been done in large $\mu_B$ limit~\cite{Pisarski:1999tv,Brown:1999aq,Brown:2000eh,Brown:1999yd,Wang:2001aq} and the expressions of the diquark condensate and the critical temperature of color superconductivity were obtained up to sub-leading order in the perturbative coupling expansion~\cite{Brown:1999aq,Wang:2001aq}. The expression of the pion condensate (more precisely, the excitation gap of the quasiparticles) was argued to be in the form $\Delta=b_\pi{|\mu_I|}|g|^{-5}e^{-{3\pi^2/(2g)}}$ with $b_\pi\approx10^4$~\cite{Son:2000xc,Son:2000by}. The actual value of $b_\pi$ is still unknown so far; so we'd like to calculate $b_\pi$ first within the perturbative framework.

\emph{The coefficient $b_\pi$.} To obtain the correct value of $b_\pi$, we need to self-consistently take the self-energy of quarks into account~\cite{Wang:2001aq}. Considering pion condensate $\pi^-(K)\neq0$ which interacts with quarks through $\bar{\psi}i\gamma_5\tau^-\pi^-(K)\psi$ where $\tau^\pm=(\tau_1\pm i\tau_2)/2$, the inverse quark propagator in flavor and energy-momentum spaces is given as
\begin{eqnarray}
G^{-1}(K)=\left(\!\begin{array}{cc}
{\slashed K}+\mu\gamma_0+\Sigma_{uu}(K)&-i\gamma_5\pi^+(K)\\
-i\gamma_5\pi^-(K)&{\slashed K}-\mu\gamma_0+\Sigma_{dd}(K)
\end{array}\!\right),
\end{eqnarray}
where we have used $\mu\equiv\mu_I/2$ to simplify the notation and set $\mu_I<0$. Then the quark propagator can be derived with the elements in flavor space given by
\begin{eqnarray}
G_{uu\!/\!dd}(K)\!\!\!&=&\!\!\!\Big(\![G_{uu\!/\!dd}^0(K)]^{-1}\!\!+\!\gamma_5\pi^{\pm}[G_{dd\!/\!uu}^0]^{-1}
\gamma_5\pi^{\mp}\!\Big)^{-1},\nonumber\\
G_{ud\!/\!du}(K)\!\!&=&\!\!iG_{uu/dd}^0(K)\gamma_5\pi^{\pm}(K)G_{dd/uu}(K),
\end{eqnarray}
where we have defined the free quark propagators with self-energy as
\begin{eqnarray}
G_{uu/dd}^0(K)\equiv[{\slashed K}\pm\mu\gamma_0+\Sigma_{uu/dd}(K)]^{-1}.
\end{eqnarray}
The self-energy can be evaluated to leading order as~\cite{Manuel:2000mk,Wang:2001aq}
\begin{eqnarray}
\Sigma_{uu/dd}(K)&=&\gamma_0{\bar g}^2(k_0\ln{M^2\over k_0^2}+i\pi|k_0|),
\end{eqnarray}
where ${\bar g}=g/(3\sqrt{2}\pi)$ and $M^2=(3\pi/4)m_g^2$ with the gluon mass $m_g^2=g^2\mu^2/(3\pi^2)$. Because the imaginary parts are suppressed by $g$ compared to the real parts and can be neglected~\cite{Wang:2001aq}, the free quark propagators can be rewritten as
\begin{eqnarray}
G_{uu/dd}^0(K)&=&[\gamma_0{\tilde{k}_0}-\gamma{\bf \cdot k}\pm\mu\gamma_0]^{-1},
\end{eqnarray}
where $\tilde{k}_0\!=\!k_0/Z(k_0)$ and $Z(k_0)\!\equiv\![1+{\bar g}^2\ln{(M^2/k_0^2)}]^{-1}$.

The gap equations for $\pi^{\pm}(K)$ can be evaluated in mean field approximation as
\begin{eqnarray}
i\gamma_5\pi^{\pm}(K)=g^2{T\over V}\sum_Q\Delta^{ab}_{\mu\nu}(K-Q)\gamma^\mu \lambda_aG_{ud/du}(Q)\gamma^\nu \lambda_b.
\end{eqnarray}
Let's focus on the gap equation for $\pi^-(K)$ and the case for $\pi^+(K)$ is similar. The color indices can be easily summed to give quardratic Casimir operator $C_2(N_c)=4/3$ and the matrix $-i\gamma_5$ can be cancelled from both sides:
\begin{eqnarray}
\pi^-\!(\!K\!)\!=\!{4g^2\over3}{T\over V}\sum_Q\Delta_{\mu\nu}\!(\!K\!-\!Q\!)\gamma^\mu G_{dd}^0(\!Q\!)\pi^-\!(\!Q\!)G_{uu}(\!Q\!)\gamma^\nu.
\end{eqnarray}
Actually, the Dirac structures of $G_{uu}(K)$ and $G_{dd}(K)$ are very simple~\cite{He:2005nk}:
\begin{eqnarray}
G_{uu\!/\!dd}(\!K\!)\!=\!{\tilde{k}_0\!\pm\!(k\!-\!\mu)\over\tilde{k}_0^2-(E_{\bf k}^-)^2}\Lambda_{\pm}(\!{\bf k}\!)\gamma_0\!+\!{\tilde{k}_0\!\mp\!(k\!+\!\mu)\over\tilde{k}_0^2-(E_{\bf k}^+)^2}\Lambda_{\mp}(\!{\bf k}\!)\gamma_0,
\end{eqnarray}
where the dispersions are $E_{\bf k}^\pm\!=\![(k\!\pm\!\mu)^2\!+\!|\pi^{-}|^2]^{1\over2}$ and the energy projectors are $\Lambda_\pm(\!{\bf k}\!)\!=\!{(1\!\pm\!\gamma_0{\gamma}\mathbf{\cdot\hat{k}})/2}$. Writing the pion condensate as $\pi^-=\sum_{s=\pm}\Lambda_s\pi^-_s$, the gap equation becomes
\begin{eqnarray}
\pi_k
\!\!=\!\!{2g^2\over3}{T\over V}\sum_Q\Delta_{\mu\nu}\!(\!K\!-\!Q\!){\pi_q\over\tilde{q}_0^2\!-\!(\!E_{\bf q}^+\!)^2}{\rm Tr}[\Lambda_-\!(\!{\bf k}\!)\gamma^\mu \Lambda_+\!(\!{\bf q}\!)\gamma^\nu].
\end{eqnarray}
Here, because $\mu<0$, it is enough to consider only $\pi_q\equiv\pi^-_-(Q)$ on the righthand side. Then, following similar discussions with Ref.~\cite{Pisarski:1999tv} and Ref.~\cite{Wang:2001aq}, we can obtain the following gap equation:
\begin{eqnarray}
\pi_k\!\simeq\!{\bar{g}^2}\!\!\!\int_0^\delta\!\!\!{\pi_qd(q\!+\!\mu)\over Z^{-2}(\tilde{E}_{\bf q}^+)\tilde{E}_{\bf q}^+}\tanh\!\Big(\!{\tilde{E}_{\bf q}^+\over2T}\!\Big)\ln\!\Big(\!{\tilde{b}^2\mu^2\over|(\!\tilde{E}_{\bf q}^+\!)^2\!-\!(\!\tilde{E}_{\bf k}^+\!)^2|}\!\Big),
\end{eqnarray}
where we have defined a constant $\tilde{b}\equiv256\pi^4|g|^{-5}$. With the redefinition $z\equiv-(2\bar{g}^2)^{-1/3}(1-2\bar{g}x)$ compared to that in Ref.~\cite{Wang:2001aq}, the new forms of Eq. (30) and Eq. (31) in Ref.~\cite{Wang:2001aq} can be obtained by just taking the combined transformations: $\bar{g}\rightarrow\bar{g}/\sqrt{2}$ and $x\rightarrow x\sqrt{2}$. Then we can immediately find from Eq.(33) in Ref.~\cite{Wang:2001aq} that
\begin{eqnarray}
\Delta=b_\pi|g|^{-5}{|\mu_I|}e^{-{\pi\over2\sqrt{2}\bar{g}}},\
b_\pi=256\pi^4e^{-{4+\pi^2\over16}}.
\end{eqnarray}
Thus, we confirm the expression of the pion condensate as a function of $g$ and $\mu_I$ given in Ref.~\cite{Son:2000xc,Son:2000by} and also find that the coefficient $b_\pi$ is less suppressed by $e^{{4+\pi^2\over16}}\approx2.38$ compared to that in color superconductor~\cite{Wang:2001aq}.

\emph{Pure gluodynamics approximation.} As shown in Ref.~\cite{Rischke:2000cn}, at low temperature, the dynamics of two-flavor color superconductivity is described by a pure $SU(2)$ gluon dynamics at large $\mu_B$. Similarly, low-temperature QCD at large $|\mu_I|$ is described by $SU(3)$ gluon dynamics in addition to the dynamics of Goldstone mode~\cite{Son:2000xc,Son:2000by,Cohen:2015soa}. By following the formalism developed in Ref.~\cite{Cohen:2015soa}, we can write down the thermodynamic potential as
\begin{eqnarray}
\Omega={\cal U}(\Phi)-{\pi^2\over90}{T^4\over \upsilon^3}+\Omega_q,
\end{eqnarray}
where $\upsilon\approx1/\sqrt{3}$ is the speed of sound and the pure gauge potential is modified from LQCD simulation by changing $T_0=270MeV$ to $\tilde{T}_0=T_0({\tilde{\Lambda}\!/\!\Lambda_{QCD}})$~\cite{Cohen:2015soa,Roessner:2006xn}:
\begin{eqnarray}
&&\!\!\!\!\!\!\!\!{{\cal U}(\Phi)\over T^4}\!=\!-{a(T)\over2}\Phi^2\!+\!b(T)\log[1\!-\!6\Phi^2\!+\!8\Phi^3\!-\!3\Phi^4],\\
&&\!\!\!\!a(T)\!=\!a_0\!+\!a_1\Big({\tilde{T}_0\over T}\Big)\!+\!a_2\Big({\tilde{T}_0\over T}\Big)^2, \ \ b(T)\!=\!b_3\Big({\tilde{T}_0\over T}\Big)^3,
\end{eqnarray}
where the modified confinement scale $\tilde{\Lambda}$ is determined by~\cite{Cohen:2015soa,Rischke:2000cn}
\begin{eqnarray}
\alpha(\Delta,\tilde\Lambda)|_{N_f=0}={\alpha(\mu_I,\Lambda)\over\sqrt{\epsilon}}={\alpha(\mu_I,\Lambda)\over
\sqrt{1+{\alpha(\mu_I,\Lambda)\mu_I^2\over72\Delta^2}}}.
\end{eqnarray}
In order to be more specific, we adopt the four-loop expression for $\alpha(\mu_I,\Lambda)$~\cite{Olive2014}.
Inspired by the Polyakov--Nambu--Jona-Lasinio model, the $\mu_I$ related quark contribution can be written in the form of free quasi-quarks:
\begin{widetext}
\begin{eqnarray}
\Omega_q&=&-2\int{d^3p\over(2\pi)^3}\sum_{s=\pm}\Big\{N_cE^s_{\bf p}+2T\ln\Big(1+3\Phi e^{-E^s_{\bf p}/T}+3\Phi e^{-2E^s_{\bf p}/T}+e^{-3E^s_{\bf p}/T}\Big)\Big\},
\end{eqnarray}
\end{widetext}
where an ultraviolet regulator is needed and can be implemented by a finite renormalization scale $M$.  Note that the temperature dependence of $\Delta$ is neglected as we are only interested in the low-temperature case $T\ll|\mu_I|$.

\begin{figure}[!htb]
\begin{center}
\includegraphics[width=8cm]{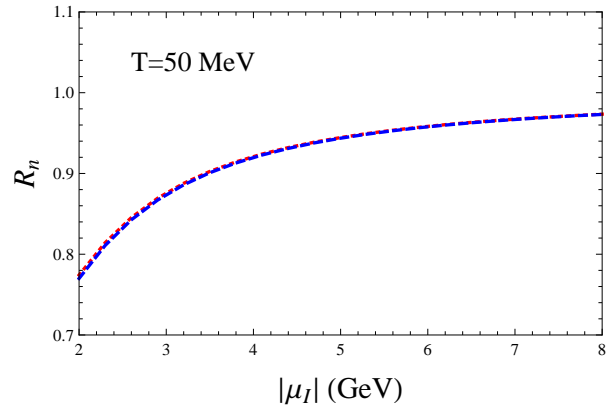}
\caption{The ratios $R_n$ of the isospin density to that in Stefan-Boltzmann limit $\Delta=0$ as a function of $|\mu_I|$ at temperature $T=50MeV$. Here, red dotted and blue dashed lines correspond to confined and deconfined phases, respectively.}\label{density}
\end{center}
\end{figure}

\begin{figure}[!htb]
\begin{center}
\includegraphics[width=8cm]{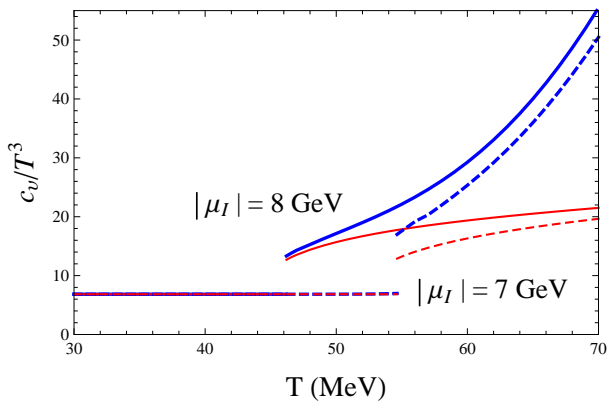}
\caption{The heat capacity $c_\upsilon$ as a function of $T$: the thick blue lines are the total contributions from quarks, gluons, and the Goldstone mode and the thin red lines are those without contributions from quarks. The solid and dashed lines correspond to $|\mu_I|=8GeV$ and $|\mu_I|=7GeV$ with deconfinement temperatures $T_d=46.19MeV$ and $T_d=54.64MeV$, respectively. The horizon lines are mainly contributed from Goldstone mode with $c_\upsilon/T^3={2\pi^2/15\upsilon^3}$ independent of $|\mu_I|$.}\label{cv}
\end{center}
\end{figure}

First, we show the ratios of the isospin density to that in the Stefan-Boltzmann limit $\Delta=0$, that is, $R_n=n_I(\Delta)/n_I(0)$  in Fig. \ref{density} for both confined ($\Phi=0$) and deconfined ($\Phi=1$) phases. As can be seen, the confinement properties have very small effect to the density at low temperature, and the system can well be described by free quark (antiquark) gas in the whole region as $R_n\approx1$, especially for $|\mu_I|>5GeV$; indicating the quarksonic nature. The heat capacity $c_\upsilon=T\partial s/\partial T|_{\mu_I}$ is explored further as shown in Fig.~\ref{cv}: Although the effective dynamics is pure gluodynamics plus Goldstone mode at low temperature~\cite{Cohen:2015soa}, the contribution to $c_\upsilon$ is dominated by the Goldstone mode. Just above the deconfinement temperature, the freed gluons start to contribute immediately and further contribution can also come from the quark (antiquark) sea. For larger temperature, it is dominated by quark part when a considerable ratio of the huge quark (antiquark) sea is realised.

\section{Summary}
In this Letter, we argue the existence of a quarksonic matter at large isospin chemical potential $|\mu_I|$ in which the quark (antiquark) Fermi seas and confinement coexist. We discuss the physical properties of the quarksonic matter in both large $N_c$ and asymptotically free limits. In large $N_c$ limit, we sketch a phase diagram in $T-|\mu_I|$ plane and argue that the quarksonic matter occupies the large $|\mu_I|$ (but starts at the order $o(N_c^0)$) and low temperature region below the deconfinement temperature $T_d$ which is independent of $|\mu_I|$. For real QCD with $N_c=3$, $T_d$ decreases with $|\mu_I|$ and thus shrinks the quarksonic matter region. In asymptotically free limit, we first obtain the coefficient $b_\pi$ which governs the pion condensate gap up to sub-leading order in the QCD coupling constant. Then, we calculate the isospin density to show that the quark part is indeed a weakly interacting gas at large $|\mu_I|$, and the heat capacity is mainly contributed from Goldstone mode at low temperature and  gluons and quarks start to contribute above the deconfinement transition. Compared to quarkyonic matter, the collapse of quarksonic matter is also caused by deconfinement transition but there are several differences: the latter is also characterized by the existence of pion superfluidity and associated Goldstone mode which can survive even above the deconfinement transition temperature.

%In principle, quarkyonic matter like phase might be found in other circumstances, such as in curved space~\cite{Flachi:2014jra} and electric field in the sense that chiral restoration happens earlier than deconfinement transition, if the parameters only directly function through quarks and chiral symmetry tends to be restored with the parameters. Furthermore, it is also a good idea to explore these exotic states by adding other chiral symmetry breaking catalysis parameters, such as magnetic field and chiral chemical potential. Our work on the effect of electric field will come out quite soon.

\emph{Acknowledgments}---
This work is supported by the Thousand Young Talents Program of China. GC and XGH are also supported by Shanghai Natural Science Foundation with Grant No. 14ZR1403000 and NSFC with Grant No. 11535012. GC is also supported by China Postdoctoral Science Foundation with Grant No. KLH1512072.

\end{document}